\documentclass{eptcs}
\usepackage[utf8]{inputenc}
\usepackage{underscore}
\usepackage{b}

\usepackage{graphicx}

\title{Interfacing Automatic Proof Agents in Atelier~B:\\ Introducing ``iapa''\thanks{This work is partly supported by the BWare (ANR-12-INSE-0010, \url{http://bware.lri.fr/}) project of the French national research organization (ANR).}}

\author{Lilian Burdy
\institute{ClearSy System Engineering, Aix-en-Provence, France} 
\email{lilian.burdy@clearsy.com}
\and 
David D\'{e}harbe\thanks{On leave from Universidade Federal do Rio Grande do Norte.}
\email{david.deharbe@clearsy.com}
\institute{ClearSy System Engineering, Aix-en-Provence, France} 
\and
\'{E}tienne Prun
\institute{ClearSy System Engineering, Aix-en-Provence, France} 
\email{etienne.prun@clearsy.com}
}
\def\titlerunning{Interfacing Automatic Proof Agents in Atelier~B}
\def\authorrunning{L. Burdy, D. D\'{e}harbe \& \'{E}. Prun}

\begin{document}

\maketitle

\begin{abstract}
The application of automatic theorem provers to discharge proof obligations
is necessary to apply formal methods in an efficient manner. Tools supporting
formal methods, such as Atelier~B, generate proof obligations fully automatically.
Consequently, such proof obligations are often cluttered with information that is
irrelevant to establish their validity. 

We present iapa, an ``Interface to Automatic Proof Agents'', a new tool that is
being integrated to Atelier~B, through which the user will access proof 
obligations, apply operations to simplify these proof obligations, and then
dispatch the resulting, simplified, proof obligations to a portfolio of 
automatic theorem provers.
\end{abstract}

\section{Introduction}

Historically, the B Method\cite{bbook} was introduced in the late 80’s to design provably safe software. Promoted and supported by RATP, the B method and Atelier~B, the tool implementing it, have been successfully applied to the industry of transportation leading to a worldwide implementation of the B technology for safety critical software, mainly as automatic train controllers for subways.

The development of such controllers corresponds to a "big" industrial project. To give an idea of the size of such a development, a train controller is composed of different software components communicating together. Taking from a real example, the size of the critical parts of a real-life controller is around 500.000 lines of B which give after translation 300.000 lines of Ada and 160.000 generated proof obligations. The proof, already mainly automated, of those proof obligations is a substantial part of the development cost for such projects.
Limiting these costs by using more efficient provers, or by using more efficiently provers is a real concern in industry.

Atelier~B comes with two provers developed in the 90's. Recently the ProB model checker~\cite{leuschel2003prob} has been added as a prover that can be called during an interactive proof session. The BWare project\cite{Bware} \cite{DDM14} aims to provide a mechanized framework to apply automated theorem provers, such as first order provers and SMT (Satisfiability Modulo Theories) solvers on proof obligations coming from the development of industrial applications using the B method. This approach produces proof obligations in the format of Why3~\cite{bobot2011why3}, which is used then responsible for calling different automatic theorem provers with the adequate input, interpreting their output and synthesizing a verification result.
%Finally this framework will be integrated to Atelier~B.

SMT provers are routinely used by any tool that has to deal with a logic-based verification task. Notably, they have already been added as plugins in Rodin\cite{Deharbe:2012:SSR:2345997.2346015}, another IDE for Event-B, which is a formal method closely related to the B method. Concerning BWare, promising first results\cite{mentre:hal-00681781} \cite{Conchon2014} have already been published.

We present here the integration of such automated provers in Atelier~B considering the specificities of proof obligations produced by industrial software developments, described in Section~\ref{sec:po}. The basic elements of this integration are the Why3-based bridge to automated theorem provers developed in the BWare project and a new approach for efficiently selecting the relevant parts of a proof obligation. The principles of this latter aspect are presented in Section~\ref{sec:core}. The implementation of this functionality in a graphical user interface is then presented in Section~\ref{sec:gui}.

\section{Presenting the B method}
The development of a project using the B method comprises two activities that are closely linked: writing formal texts and proving these same texts.

The writing activity consists in writing the specifications for \emph{abstract machines} using a high level mathematical formalism. In this way, a B specification comprises data (that may be expressed among other ways using integers, Boolean values, sets, relations, functions or sequences), \emph{invariant properties} that relate to the data (expressed using first order logic), and finally services that describe the initialization and possible evolutions of the data (data transformations being expressed as substitutions). The proof activity for a B specification comprises performing a number of demonstrations in order to prove the establishment and conservation of invariant properties in the specification (e.g. it is necessary to prove that a service call retains the invariant properties). The generation of such proof obligations is based essentially on the transformation of predicates using substitutions.

The development of an abstract machine continues %using an extension to the write activity 
during the successive \emph{refinement} steps. Refining a specification consists in reformulating it so as to provide more and more concrete solutions, but also to enrich it. The proof activity relating to refinements performs a number of static checks and demonstrations in order to prove that the refinement is a valid reformulation of the specification. 

\section{Proof obligations}
\label{sec:po}

Verifying program by proving verification conditions (called proof obligations here) leads to deal with huge lemmas. This problem is not specific to the B Method, indeed the same observation is done, for example, in \cite{ch07:bl} for verification conditions issued from C program verification.

Concerning the B Method\cite{bbook}, Figure \ref{fig:po} shows a proof obligation template. One can see that hypotheses are collected in many clauses of many components. As an example, let us consider the declaration of constants and their properties: as the B Method is a modular software development method, constants are usually declared in some specific components. When a function needs to use a constant, the component in which the constant is declared is \emph{seen}\footnote{The SEES link is used to reference within a component, an abstract machine instance, to access its constituents (sets, constants and variables) without modifying them.} and with the needed constant come all the properties of all the constants of the component. So all the proof obligations will have as hypotheses the properties of all the constants of the seen component even if only few of them are relevant for the proof. 

\begin{figure}[!th]
\[
\begin{array}{l}
\mbox{\emph{``Machine parameter 
constraints''}} \band \\

\mbox{\emph{``Properties of 
constants of
    previous refinements''}}  \band \\

\mbox{\emph{``Properties of refinement constants''}} \band 
\\

\mbox{\emph{``Properties of constants of components 
seen''}} \band \\

\mbox{\emph{``Properties of constants of 
components included''}} \band \\

\mbox{\emph{``Invariants and
  assertions of included components''}} \band
\\

\mbox{\emph{``Invariants and assertions of components seen''}} \band
\\

\mbox{\emph{``Invariants and assertions of the vertical 
development''}} \band \\

\mbox{\emph{``Precondition of the abstract 
operation''}}\\

\!\!\Rightarrow   \\

\mbox{\emph{``Precondition of the refinement 
operation''}} \band  \\

  \mbox{\emph{``Refinement operation applied to the
      negation of the specified operation applied to the negation }}
 \\

  \mbox{\quad\emph{ of the invariant''}}
\end{array}
\]
\caption{Refinement proof obligation template}
\label{fig:po}
\end{figure}

The incremental approach to refinement is another source of growth in the size of the proof obligations. Indeed, in each new refinement, all the proof obligations will include the contexts from all the components that come before in the refinement chain. 

In the real project described previously, the average number of hypotheses for a lemma is around 2000 formulas. Some proof obligations can contain more that 4000 hypotheses. Of course, not all of them are necessary in the proof of each goal.
To use efficiently automated provers on such lemmas, we argue that it is necessary to filter relevant hypotheses. This is the motivation for the development of an ``Interface to Automatic Proof Agents'' (IAPA), giving the users of Atelier~B the means to build scripts constructing a mini-lemma from a generated proof obligation, and to submit such mini-lemmas to the provers.

\section{Core functionality in iapa \label{sec:core}}

Atelier~B already contains an interface to discharge proof
obligations: the interactive prover, where users spend most of their
time. Some functionalities in iapa are similar to those found in the
interactive prover and will be familiar to the users.

The iapa tool is invoked within Atelier~B on a given component, once
the proof obligations of that component have been generated (as with
the interactive prover). In Atelier~B, all the proof obligations of a
component are available in a single file, and can be either in a
legacy format called the ``theory language'' or in a XML-based
format. Only the latter contains typing annotations and this
format has been chosen in the BWare initiative as the basis for
interfacing with the Why3 platform. Thus, iapa reads the proof
obligations of a given component from the XML-based file.

In the iapa interface, proof obligations are grouped according to
their origin in the corresponding B component (assertions,
initialization, operations). Navigating through these proof
obligations is a first core functionality available in iapa, and its
principles mimick those of the interactive prover. Regarding this
aspect, one important difference with respect to the interactive
prover is that well-definedness proof obligations are presented
together with those proof obligations instead of in a separate
project.

Formally, a proof obligation is a pair $(\Gamma, \varphi)$, where
$\Gamma$ is a set of hypotheses and $\varphi$ is the goal.  The main
goal of iapa is to assist the user in selecting the relevant
information in the current proof obligation. This consists
in building a new proof obligation $(\Gamma', \varphi)$ such that
$\Gamma' \subseteq \Gamma$. In iapa, the new, simplified, proof
obligation is termed the \emph{lemma}. The interface also contains
means to submit lemmas to a portfolio of automatic theorem provers.

One notable requirement of iapa is that the steps leading to the
construction of a lemma for a given proof obligation can be applied
automatically to other proof obligations. This is achieved by means of
two kinds of entities: \emph{contexts} and \emph{lexicons}, that the
user has to manipulate and combine in order to build a lemma.

A context $\gamma \subseteq \Gamma$ is a set of hypotheses that
originates from the proof obligation $(\Gamma, \varphi)$. When a proof
obligation is loaded in iapa, a number of contexts are pre-defined:
\begin{itemize}
\item \textsf{all} contains all the hypotheses;
\item \textsf{local} contains all the hypotheses that are local in the
  proof obligation, i.e. hypotheses stemming from conditions found in 
  the operation corresponding to the proof obligation;
\item \textsf{global} contains all but the local hypotheses;
\item a number of contexts that correspond to the different
  sections in a B component (properties, invariants, etc.);
\item \textsf{B definitions} contains hypotheses on pre-defined sets
  such as implementable integers.
\end{itemize}

A lexicon $l$ is a set of free identifiers of the original proof
obligations. Assuming $\mathsf{fv}$ returns the set of free
identifiers in a formula, then $l \subseteq \bigcup \{
\mathsf{fv}(\psi) \mid \psi \in \Gamma \cup \{ \varphi \}
\}$. Initially, there is a single pre-defined lexicon, named
\textsf{goal}, and containing $\mathsf{fv}(\varphi)$ (the free
identifiers in the goal).

At any time, the state of iapa contains the following elements:
\begin{itemize}
\item $(\Gamma, \varphi)$ the current proof obligation;
\item $\mathcal{C}$: a set of contexts ($\forall x \in \mathcal{C}, x \subseteq \Gamma$);
\item $\mathcal{L}$: a set of lexicons ($\forall l \in \mathcal{L}, l \subseteq \bigcup \{
\mathsf{fv}(\psi) \mid \psi \in \Gamma \cup \{ \varphi \}
\}$);
\item $c$: a current context ($c \in \mathcal{C}$);
\item $l$: a current lexicon ($l \in \mathcal{L}$);
\item $S$: a set of selected hypotheses ($S \subseteq \Gamma$).
\end{itemize}
The pre-defined values for $\mathcal{C}$ and $\mathcal{L}$ are as
described previously, those of $c$, $l$ and $S$ are \textsf{local},
\textsf{goal}, and $\emptyset$, respectively. Then the commands on
contexts and lexicons currently implemented in iapa are the following:
\begin{itemize}
\item \textsf{ah} adds the hypotheses in the current context to 
the set of selected hypotheses;
\item \textsf{dh} removes the hypotheses in the current context from
  the set of selected hypotheses;
\item \textsf{chctx(ctx)} sets the current context to $\textsf{ctx}$;
\item \textsf{chlex(lex)} sets the current lexicon to $\textsf{lex}$;
\item \textsf{mklex} creates a new lexicon with the free identifiers 
of the current context;
\item \textsf{mklex(i1,..., in)} creates a new lexicon with
the given identifiers;
\item \textsf{mkctx(Some)} creates a new context containing the hypotheses
in the current context that have at least one free identifier in the current
lexicon;
\item \textsf{mkctx(All)} creates a new context containing the hypotheses
in the current context such that their free identifiers includes the current
lexicon;
\item \textsf{mkctx(h1,...,hn)} creates a new context containing the given
hypotheses.
\end{itemize}
The condition and effect of the execution of these commands are
summarized in Table~\ref{tab:semantics}, and some illustrative
scripts are presented in Table~\ref{tab:scripts}. The following section
presents the iapa interface, including how the user can play such commands.

\begin{table}
\begin{center}
\begin{tabular}{l|l|l}
\hline
command & effect & condition \\
\hline
\hline
\textsf{ah} & $S := S \cup c $ & \\
\textsf{dh} & $S := S \setminus c$ & \\
\textsf{chctx(ctx)} & $c := \textsf{ctx}$ & $\textsf{ctx} \in \mathcal{C}$ \\
\textsf{chctx(lex)} & $l := \textsf{lex}$ & $\textsf{lex} \in \mathcal{L}$ \\
\textsf{mklex} & $\mathcal{L} := \mathcal{L} \cup \{ \mathsf{fv}(c) \}$ & $\mathsf{fv}(c) \neq \emptyset$ \\
\textsf{mklex(i1,...,in)} &  $\mathcal{L} := \mathcal{L} \cup \{\{\mathsf{i1},\cdots,\mathsf{in}\}\}$ & $\mathsf{i1} \in l \cdots \mathsf{in} \in l$ \\
\textsf{mkctx(Some)} & $\mathcal{C} := \mathcal{C} \cup \{ \{ h | h \in c \land \mathsf{fv}(h) \cap l \neq \emptyset \} \}$  & $\{ h | h \in c \land \mathsf{fv}(h) \cap l \neq \emptyset \} \neq \emptyset$ \\
\textsf{mkctx(All)} & $\mathcal{C} := \mathcal{C} \cup \{ \{ h | h \in c \land l \subseteq \mathsf{fv}(h) \} \}$ & $\{ h | h \in c \land l \subseteq \mathsf{fv}(h) \} \neq \emptyset$ \\
\textsf{mkctx(h1,...,hn)} &  $\mathcal{C} := \mathcal{C} \cup \{\{\mathsf{h1},\cdots,\mathsf{hn}\}\}$ & $\mathsf{h1} \in c \cdots \mathsf{hn} \in c$ \\
\hline
\end{tabular}
\end{center}
\caption{Formalization of iapa commands. \label{tab:semantics}}
\end{table}

\begin{table}
\begin{center}
\begin{tabular}{p{.43\textwidth}|p{.5\textwidth}}
\hline
script & description \\
\hline
\hline
\textsf{ah} & builds the lemma containing only local hypotheses \\
\hline
\textsf{chctx(all) \& ah} & builds the lemma identical to proof obligation \\
\hline
\textsf{mklex \& chctx(all) \& mkctx(Some) \& ah} & builds the lemma with hypotheses containing an identifier in the local hypotheses \\
\hline
\end{tabular}
\end{center}
\caption{Example iapa scripts (\& being used as command separator). \label{tab:scripts}}
\end{table}

\section{The iapa tool \label{sec:gui}}

\begin{figure}
\begin{center}
\includegraphics[height=.65\textheight]{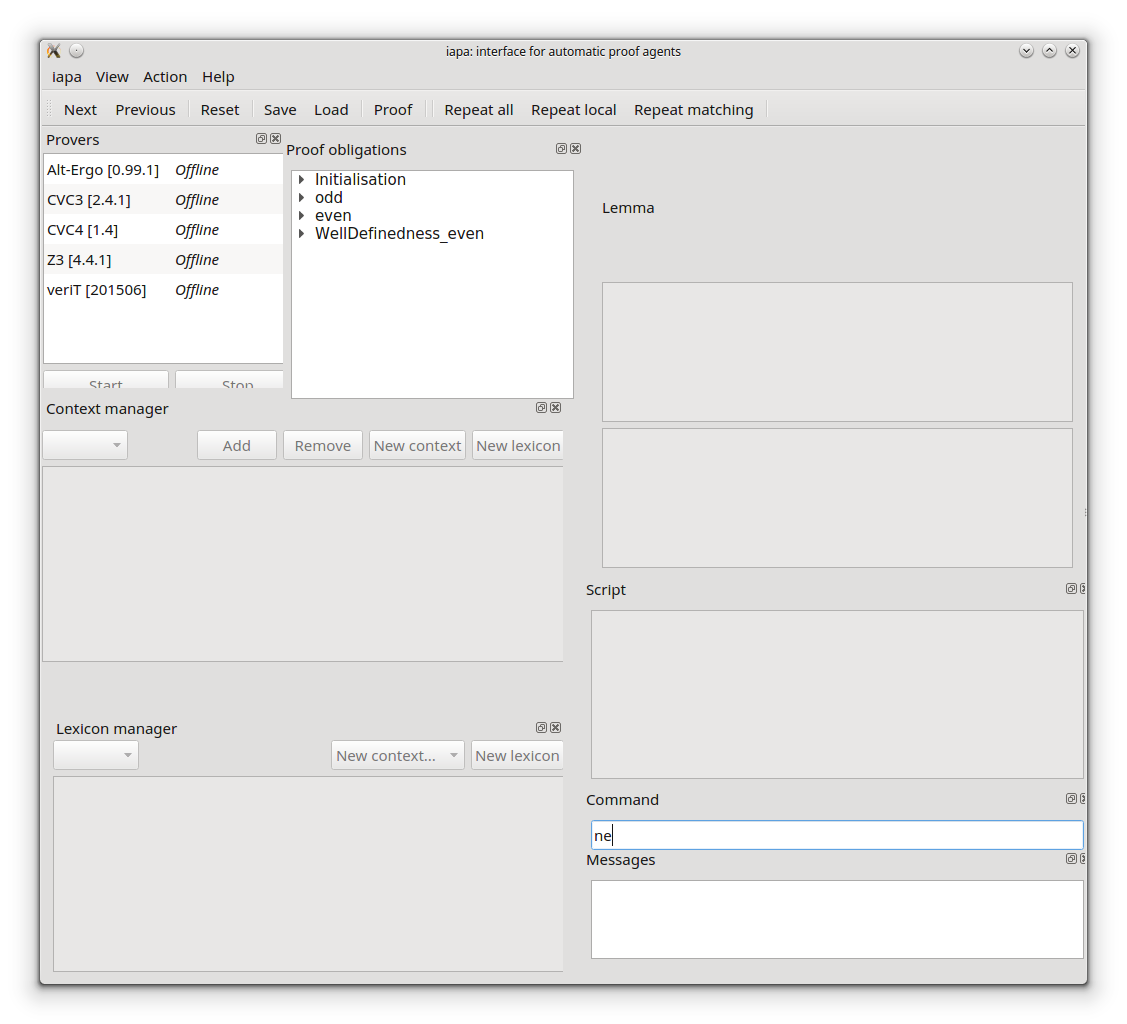}
\end{center}
\caption{Initial contents of the iapa window.\label{fig:initial}}
\end{figure}

The core functionality is presented in a graphical user interface that
has to be launched from Atelier~B's main window on a given
component. The functionalities are offered both by textual and
point-and-click means. Figure~\ref{fig:initial} contains a screenshot
of the initial contents of the window. At that point, two views are
populated: \emph{Provers} and \emph{Proof obligations}. The contents
of the latter is found in Atelier~B's database. The contents of the
former is obtained by querying the automatic provers currently
installed. Since, at the moment, the access to these provers is
realized through Why3, this information is found automatically using
Why3 and its auto-configuration facilities. Besides the menu and the
tool bar found at the top of the window, the \emph{Command} section
contains a widget containing a command-line interface to iapa. Here,
the user has already typed the command \textsf{ne}, which, when
executed, will open the next proof obligation.

\begin{figure}
\begin{center}
\includegraphics[height=.65\textheight]{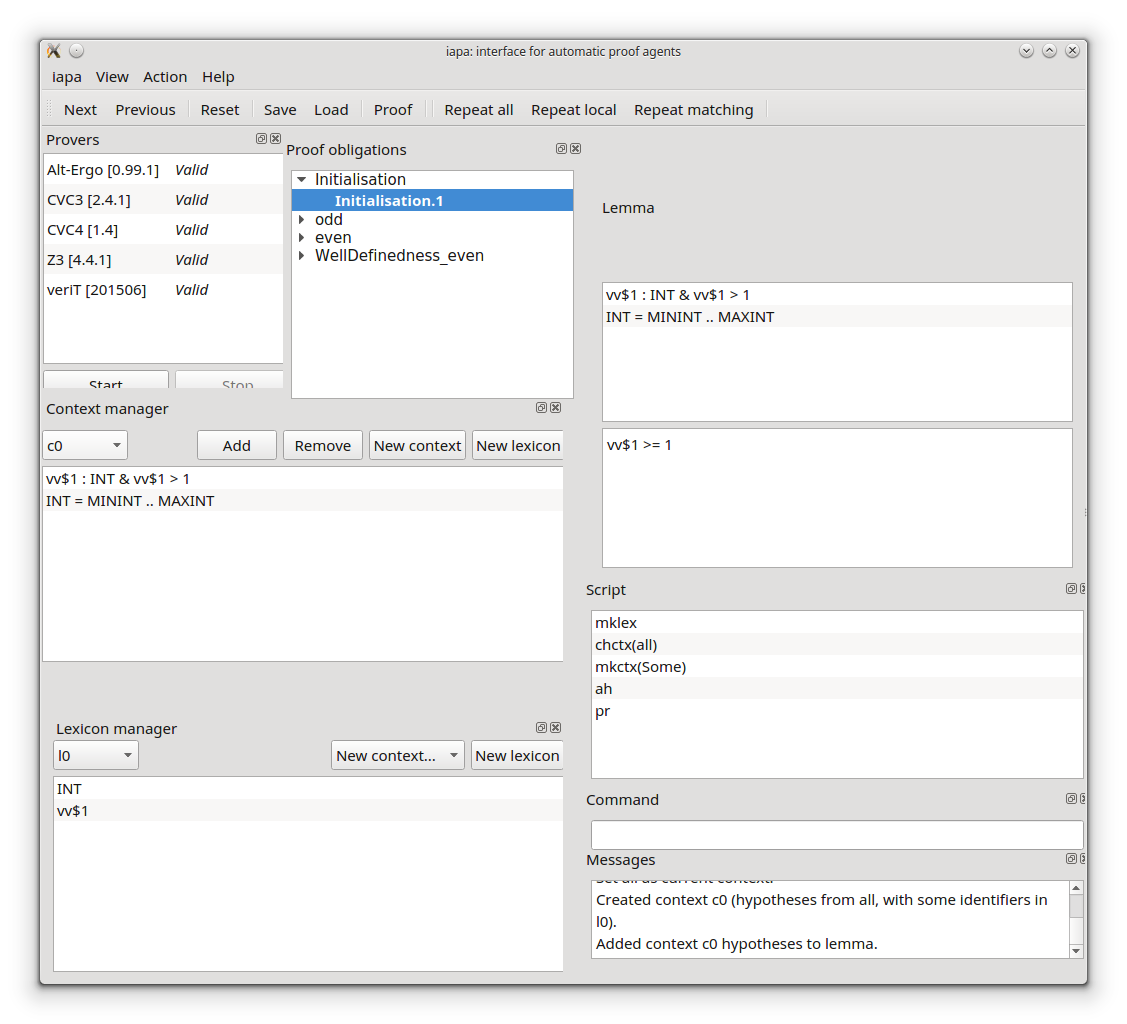}
\end{center}
\caption{Contents of the iapa window during a session.\label{fig:final}}
\end{figure}

Opening a proof obligation results in filling the \emph{Goal},
\emph{Context manager} and \emph{Lexicon manager} sections, and
enables actions corresponding to the core iapa functionalities.
Figure~\ref{fig:final} shows the contents of the iapa window after the
user has managed to complete a proof after having selected some
hypotheses in the context (using here one of the scripts presented in
Table~\ref{tab:scripts}) and started the provers on the resulting
lemma with the \textsf{pr} command. Scenarios where the proof
obligation is not found valid correspond to the following cases: the
original proof obligation is not valid, the user did not select a
relevant subset of hypotheses, or the tools did not find the proof.

In any case, the steps realized by the user are saved and displayed in
the \emph{Script} section. Also the \emph{Messages} section is
dedicated to the display of feed-back information.

%% \begin{figure}
%% \begin{center}
%% \includegraphics[width=.9\textwidth]{images/final.png}
%% \end{center}
%% \caption{Contents of the iapa window during a session.\label{fig:final}}
%% \end{figure}

\section{Conclusions and future work}

This paper presents an on-going work to reduce the cost of discharging
proof obligations when applying formal methods in an industrial
environment. This work is embodied in \textsf{iapa}, a prototype for
an extension of Atelier~B aiming at both integrating additional proof
engines and offering hypotheses selection facilities to the user. 

The effectiveness of the approach will be assessed through a
systematic evaluation on a representative set of industrial
projects. The results of this evaluation will decide whether
\textsf{iapa} is eventually deployed together with the distributions
of Atelier~B. We also plan to improve the usability of \textsf{iapa},
by adding hypotheses selection criteria based on formula patterns, and
also taking user feed-back into account.

Certification is another important aspect of tooling in an industrial
setting for safety-critical systems. The historical provers in
Atelier~B have been certified but certifying new tools is costly. We
forecast some solutions to address this issue in iapa. First, since
some automated theorem provers are proof-producing, we envision using
the proofs thus produced to build proofs that can be played by the
certified provers. Second, since redundancy is also a mean to achieve
desired safety levels, a second tool chain could be developed. It
would bypass Why3 and generate proof obligations directly in the input
language of the automatic provers. An approach similar
to~\cite{Deharbe:2012:SSR:2345997.2346015}, targetting the SMT-LIB
format~\cite{BarFT-RR-15} is a good candidate.

\bibliographystyle{eptcs}
\bibliography{iapa}

\begin{thebibliography}{10}
\providecommand{\bibitemdeclare}[2]{}
\providecommand{\surnamestart}{}
\providecommand{\surnameend}{}
\providecommand{\urlprefix}{Available at }
\providecommand{\url}[1]{\texttt{#1}}
\providecommand{\href}[2]{\texttt{#2}}
\providecommand{\urlalt}[2]{\href{#1}{#2}}
\providecommand{\doi}[1]{doi:\urlalt{http://dx.doi.org/#1}{#1}}
\providecommand{\bibinfo}[2]{#2}

\bibitemdeclare{book}{bbook}
\bibitem{bbook}
\bibinfo{author}{Jean{-}Raymond \surnamestart Abrial\surnameend}
  (\bibinfo{year}{2005}): \emph{\bibinfo{title}{The B-book - assigning programs
  to meanings}}.
\newblock \bibinfo{publisher}{Cambridge University Press}.

\bibitemdeclare{techreport}{BarFT-RR-15}
\bibitem{BarFT-RR-15}
\bibinfo{author}{Clark \surnamestart Barrett\surnameend},
  \bibinfo{author}{Pascal \surnamestart Fontaine\surnameend} \&
  \bibinfo{author}{Cesare \surnamestart Tinelli\surnameend}
  (\bibinfo{year}{2015}): \emph{\bibinfo{title}{{The SMT-LIB Standard: Version
  2.5}}}.
\newblock \bibinfo{type}{Technical Report}, \bibinfo{institution}{Department of
  Computer Science, The University of Iowa}.
\newblock \bibinfo{note}{Available at {\tt www.SMT-LIB.org}}.

\bibitemdeclare{inproceedings}{bobot2011why3}
\bibitem{bobot2011why3}
\bibinfo{author}{Fran{\c{c}}ois \surnamestart Bobot\surnameend},
  \bibinfo{author}{Jean-Christophe \surnamestart Filli{\^a}tre\surnameend},
  \bibinfo{author}{Claude \surnamestart March{\'e}\surnameend} \&
  \bibinfo{author}{Andrei \surnamestart Paskevich\surnameend}
  (\bibinfo{year}{2011}): \emph{\bibinfo{title}{Why3: Shepherd your herd of
  provers}}.
\newblock In: {\sl \bibinfo{booktitle}{Boogie 2011: First International
  Workshop on Intermediate Verification Languages}}, pp.
  \bibinfo{pages}{53--64}.

\bibitemdeclare{inproceedings}{Conchon2014}
\bibitem{Conchon2014}
\bibinfo{author}{Sylvain \surnamestart Conchon\surnameend} \&
  \bibinfo{author}{Mohamed \surnamestart Iguernelala\surnameend}
  (\bibinfo{year}{2014}): \emph{\bibinfo{title}{Tuning the Alt-Ergo SMT Solver
  for B Proof Obligations}}.
\newblock In \bibinfo{editor}{Yamine \surnamestart Ait~Ameur\surnameend} \&
  \bibinfo{editor}{Klaus-Dieter \surnamestart Schewe\surnameend}, editors: {\sl
  \bibinfo{booktitle}{Abstract State Machines, Alloy, B, TLA, VDM, and Z: 4th
  International Conference, ABZ 2014, Toulouse, France, June 2-6, 2014.
  Proceedings}}, \bibinfo{publisher}{Springer Berlin Heidelberg},
  \bibinfo{address}{Berlin, Heidelberg}, pp. \bibinfo{pages}{294--297},
  \doi{10.1007/978-3-662-43652-3_27}.

\bibitemdeclare{inproceedings}{ch07:bl}
\bibitem{ch07:bl}
\bibinfo{author}{Jean-Fran\c{c}ois \surnamestart Couchot\surnameend} \&
  \bibinfo{author}{Thierry \surnamestart Hubert\surnameend} (\bibinfo{year}{2007}):
  \emph{\bibinfo{title}{A Graph-based Strategy for the Selection of
  Hypotheses}}.
\newblock In: {\sl \bibinfo{booktitle}{FTP'07, Int. Workshop on First-Order
  Theorem Proving}}, \bibinfo{address}{Liverpool, UK}.

\bibitemdeclare{inproceedings}{Deharbe:2012:SSR:2345997.2346015}
\bibitem{Deharbe:2012:SSR:2345997.2346015}
\bibinfo{author}{David \surnamestart D{\'e}harbe\surnameend},
  \bibinfo{author}{Pascal \surnamestart Fontaine\surnameend},
  \bibinfo{author}{Yoann \surnamestart Guyot\surnameend} \&
  \bibinfo{author}{Laurent \surnamestart Voisin\surnameend}
  (\bibinfo{year}{2012}): \emph{\bibinfo{title}{SMT Solvers for Rodin}}.
\newblock In: {\sl \bibinfo{booktitle}{Proceedings of the Third International
  Conference on Abstract State Machines, Alloy, B, VDM, and Z}},
  \bibinfo{series}{ABZ'12}, \bibinfo{publisher}{Springer-Verlag},
  \bibinfo{address}{Berlin, Heidelberg}, pp. \bibinfo{pages}{194--207},
  \doi{10.1007/978-3-642-30885-7_14}.

\bibitemdeclare{inproceedings}{DDM14}
\bibitem{DDM14}
\bibinfo{author}{David \surnamestart Delahaye\surnameend},
  \bibinfo{author}{Catherine \surnamestart Dubois\surnameend},
  \bibinfo{author}{Claude \surnamestart Marché\surnameend} \&
  \bibinfo{author}{David \surnamestart Mentré\surnameend}
  (\bibinfo{year}{2014}): \emph{\bibinfo{title}{{The BWare Project: Building a
  Proof Platform for the Automated Verification of B Proof Obligations}}}.
\newblock In: {\sl \bibinfo{booktitle}{{Abstract State Machines, Alloy, B, VDM,
  and Z (ABZ)}}}, pp.~\bibinfo{pages}{--}, \doi{10.1007/978-3-662-43652-3_26}.

\bibitemdeclare{inproceedings}{leuschel2003prob}
\bibitem{leuschel2003prob}
\bibinfo{author}{Michael \surnamestart Leuschel\surnameend} \&
  \bibinfo{author}{Michael \surnamestart Butler\surnameend}
  (\bibinfo{year}{2003}): \emph{\bibinfo{title}{ProB: A Model Checker for B}}.
\newblock In \bibinfo{editor}{Keijiro \surnamestart Araki\surnameend},
  \bibinfo{editor}{Stefania \surnamestart Gnesi\surnameend} \&
  \bibinfo{editor}{Dino \surnamestart Mandrioli\surnameend}, editors: {\sl
  \bibinfo{booktitle}{FME 2003: Formal Methods: International Symposium of
  Formal Methods Europe, Pisa, Italy, September 8-14, 2003. Proceedings}},
  \bibinfo{publisher}{Springer Berlin Heidelberg}, \bibinfo{address}{Berlin,
  Heidelberg}, pp. \bibinfo{pages}{855--874},
  \doi{10.1007/978-3-540-45236-2_46}.

\bibitemdeclare{inproceedings}{mentre:hal-00681781}
\bibitem{mentre:hal-00681781}
\bibinfo{author}{David \surnamestart Mentr{\'e}\surnameend},
  \bibinfo{author}{Claude \surnamestart March{\'e}\surnameend},
  \bibinfo{author}{Jean-Christophe \surnamestart Filli{\^a}tre\surnameend} \&
  \bibinfo{author}{Masashi \surnamestart Asuka\surnameend}
  (\bibinfo{year}{2012}): \emph{\bibinfo{title}{{Discharging Proof Obligations
  from Atelier B using Multiple Automated Prover}s}}.
\newblock In \bibinfo{editor}{Steve \surnamestart Reeves\surnameend} \&
  \bibinfo{editor}{Elvinia \surnamestart Riccobene\surnameend}, editors: {\sl
  \bibinfo{booktitle}{{ABZ - 3rd International Conference on Abstract State
  Machines, Alloy, B and Z}}}, {\sl \bibinfo{series}{Lecture Notes in Computer
  Science}} \bibinfo{volume}{7316}, \bibinfo{publisher}{{Springer}},
  \bibinfo{address}{Pisa, Italy}, pp. \bibinfo{pages}{238--251},
  \doi{10.1007/978-3-642-30885-7_17}.
\newblock \urlprefix\url{https://hal.inria.fr/hal-00681781}.

\bibitemdeclare{misc}{Bware}
\bibitem{Bware}
\bibinfo{author}{BWare \surnamestart team\surnameend} (\bibinfo{year}{2012}):
  \emph{\bibinfo{title}{The {BWare Project}}}.
\newblock \urlprefix\url{http://bware.lri.fr/}.

\end{thebibliography}

\end{document}